# Treating symptoms or root causes: How does information about causal mechanisms affect interventions?

Romy Müller

*Faculty of Psychology, Chair of Engineering Psychology and Applied Cognitive Research, TUD Dresden University of Technology, Dresden, Germany*

Corresponding author:

Romy Müller
Chair of Engineering Psychology and Applied Cognitive Research
Technische Universität Dresden
Helmholtzstraße 10, 01069 Dresden, Germany
Email: romy.mueller@tu-dresden.de
Phone: +49 351 46335330
ORCID: 0000-0003-4750-7952

## Abstract

When deciding how to solve complex problems, it seems important not only to know whether an intervention is helpful but also to understand why. Therefore, the present study investigated whether explicit information about causal mechanisms enables people to distinguish between multiple interventions. It was hypothesised that mechanism information helps them appreciate indirect interventions that treat the root causes of a problem instead of just fixing its symptoms. This was investigated in an experimental hoof trimming scenario in which participants evaluated various interventions. To do so, they received causal diagrams with different types of causal information and levels of mechanistic detail. While detailed mechanism information and its embedding in the context of other influences made participants less sceptical towards indirect interventions, the effects were quite small. Moreover, it did not mitigate participants' robust preference for interventions that only fix a problem's symptoms. Taken together, the findings suggest that in order to help people choose sustainable interventions, it is not sufficient to make information about causal mechanisms available.

# 1. Introduction

Imagine your car is overheating and you notice oil in the coolant reservoir. Would it be a better solution to flush the cooling system or to exchange a deformed head gasket? There are many ways to intervene in a complex system, but not all of them are equally useful. Some interventions actually treat the root causes of a problem and thus make sure that it does not reappear. Conversely, others treat symptoms and thus only work in the short term. Yet others have no impact on the problem at all, even aggravate it, or give rise to unintended side effects. Accordingly, successful problem solvers must be able to critically evaluate potential interventions. To this end, it often is necessary to understand the underlying causal mechanisms that make an intervention work. In the present study, a mechanism is defined in line with Ahn et al. (1995, p.309): "The mechanism information in causal reasoning specifies through which processes the event must have occurred (i.e., how a factor led to the consequence) by using vocabularies describing entities that are not presented in the event descriptions." For instance, in the car example, the head gasket separates the engine block from the cylinder head. When it gets deformed, oil and coolant can mix, so that oil will circulate in the coolant system and enter the coolant reservoir. Exchanging a deformed head gasket blocks this problematic mechanism chain. An intervention that starts even earlier in the mechanism chain would make sure that the head gasket does not get deformed in the first place. Conversely, flushing the cooling system does not do anything about the problematic mechanism chain but only fixes its symptoms.

As illustrated by this example, in order to distinguish between more and less sustainable interventions, problem solvers need to know how these interventions relate to the problem at hand. Given that people often have insufficient knowledge about causal relations, it therefore seems promising to make these relations explicit. This can be done in the form of causal diagrams – directed graphs that connect nodes with arrows to show how variables affect other variables (Pearl & Mackenzie, 2018). Yet, recent research casts doubt on whether people make adequate use of causal diagrams when choosing interventions (Kleinberg, 2024; Kleinberg et al., 2025; Kleinberg & Marsh, 2023; Korshakova et al., 2023; Zheng et al., 2020). However, the interpretation of these findings is complicated by the methodological particularities of previous studies. One could argue that they did not give causal diagrams enough of a chance to be useful. The success of causal diagrams likely depends on constraints such as whether the interventions are plausible options, whether an understanding of causal mechanisms is really needed to solve the task, and whether the causal information is designed to enhance this understanding.

With these constraints in mind, the present study asked how decisions about interventions are affected by causal diagrams. More specifically, the central research question was whether information about causal mechanisms improves a fine-grained distinction between multiple helpful interventions that differ in their sustainability. Participants had to rate and choose interventions in an unfamiliar, physical domain (i.e., equine hoof trimming). Most of these interventions had a desirable impact on the problem but systematically varied in terms of how early in the causal chain they intervened and thus how sustainable their outcome was. To investigate whether and how such interventions become distinguishable with different kinds of causal information, it was manipulated whether causal diagrams were available, whether they provided information about mechanisms, how detailed these mechanism chains were, and whether they were embedded in the context of other variables. Before reporting the study, it will be discussed why causal reasoning is challenging, how it changes with causal information, how the available interventions determine what information is needed, and what features of the information should thus be considered in an empirical investigation.

# 2. Theoretical background

## 2.1 Causal reasoning is challenging

Despite acting in a complex world, people often do not accurately represent the causal mechanisms of even simple physical systems (Lawson, 2006). Such incorrect ideas about mechanisms can impair causal reasoning (Park & Sloman, 2013). This goes hand in hand with various reasoning biases. For instance, people tend to assume linear, unidirectional cause-effects relations rather than complex system interactions (White, 1997, 2000) and they often represent causal knowledge as isolated chunks rather than interconnected networks (Johnson & Ahn, 2015). They ignore alternative causes (Fernbach et al., 2011) and have trouble recognising causal relations when causes and effects are either dissimilar (Dündar-Coecke et al., 2022; LeBoeuf & Norton, 2012; White, 2009) or temporally delayed (Lagnado & Speekenbrink, 2010).

These cognitive biases are not problematic in most everyday life situations. In fact, we get by surprisingly well with very little knowledge and understanding (Sloman & Fernbach, 2017). Even more so, incomplete representations of causal relations can be considered a rational way of utilising limited resources in a complex world (Icard & Goodman, 2015). However, a different situation arises when problems occur and we have to decide how to solve them. While most people can drive a car without understanding the physical mechanism of how it works, they will not be able to repair a car without such knowledge. When being confronted with a problem, people choose interventions according to their individual causal models (de Kwaadsteniet & Hagmayer, 2018; de Kwaadsteniet et al., 2010; Hagmayer & Sloman, 2009). Thus, accurate causal models allow them to choose better interventions (Gary & Wood, 2011). Accordingly, when people ask causal questions, a large share of these questions is about interventions (Hagmayer & Engelmann, 2020). In fact, choosing optimal interventions may be one of the key functions of causal reasoning and explanation (Kirfel et al., 2024).

However, in reality, problem solvers usually do not have someone by their side who answers their causal questions in easily digestible pieces, one cause-effect relation at a time. Although large language models (LLMs) have developed some unexpected general reasoning capabilities, accurate causal reasoning and discourse still is beyond their capabilities (Kankowski et al., 2025; Kıcıman et al., 2023; Lupyan et al., 2025; Rothschild, 2025; Yiu et al., 2024). Therefore, people often have to search for information about potential solutions to a problem. If they are lucky, this search will yield models of causal relations that have been formalised in the past. Such models could represent expert knowledge about how a particular problem is affected by different interventions. This knowledge can be made available in the form of causal diagrams that present all relevant causal relations in a clearly arranged structure (Pearl & Mackenzie, 2018). Given the importance of causal knowledge for intervention choice, one would expect such diagrams to be immensely helpful. However, recent research paints a different picture.

## 2.2 Causal diagrams do not always help to solve problems

The impact of causal diagrams on intervention choice has been studied in two contexts: continuous system control and one-shot decision-making about real-world problems.

*2.2.1 Causal diagrams for system control*

Studies on system control investigated whether people can more effectively solve complex, dynamic problems when seeing how the variables in a system are causally related (Goode & Beckmann, 2010, 2016; Klostermann & Thüring, 2007; Putz-Osterloh, 1993). Typically, participants had to control a simulation of an interconnected system like a chemical process plant by changing input variables and observing the effects on output variables. Causal diagrams showed how each input affected the outputs (e.g., input 1 increases output 1, input 2 decreases output 1 but increases output 2). In some studies, causal diagrams did not help (Klostermann & Thüring, 2007; Putz-Osterloh, 1993), while in others they did under particular circumstances (Goode, 2011; Goode & Beckmann, 2010, 2016). Thus, even when people can see all relevant causal relations in a system, this does not guarantee that they can change its performance for the better. However, in most of these studies, the actual problem contents were irrelevant to the task, while only the presence and polarity of relations between abstract variables mattered. Obviously, such setups cannot tell us how causal information affects decision-making in content-specific real-world problems that critically depend on understanding the causal mechanisms.

*2.2.2 Causal diagrams for making decisions about real-world problems*

The impact of causal diagrams on decisions about real-world problems has been investigated in the context of diagnosis and intervention choice. In the context of diagnosis, causal diagrams enabled both expert clinicians and novices to seek more relevant information and make more rational inferences (de Kwaadsteniet et al., 2013). Specifically, participants had to decide which symptoms were most informative to diagnose a mental disorder. With causal diagrams, they relied on symptoms from diverse mechanisms chains instead of narrowly focusing on symptoms that are too closely related and thus less informative. Other studies investigated how causal diagrams affect the choice of interventions for real-world problems such as physical and mental health issues, time and investment management, or fundraising (Kleinberg et al., 2025; Kleinberg & Marsh, 2023; Korshakova et al., 2023; Zheng et al., 2020). For instance, participants were asked what works best to maintain a healthy body weight, while causal diagrams presented the influences (Kleinberg & Marsh, 2023). These influences differed in their polarity (i.e., increase vs. decrease) and were arranged in different causal structures (e.g., causal chain, several causes influence a common effect). Participants received a list of interventions and had to decide which is best. One option reflected a plausible intervention, while the others were rather implausible. For instance, in the case of maintaining a healthy body weight (Kleinberg & Marsh, 2023), the options were "Don't do anything, weight is genetic", "Fast forward through TV commercials", "Get takeout pizza instead of cooking" and "Add more vegetables to his weekend recipes". In these studies, causal diagrams often did not lead to better intervention choices compared to no causal information. Notably, participants did not strictly adhere to the information provided by the diagrams (Kleinberg et al., 2025). Instead, they brought in aspects from previous knowledge and based their reasoning on information that was not covered but still relevant (e.g., feasibility and side effects of interventions, personal beliefs and preferences).

*2.2.3 Summary: Impacts of causal diagrams on intervention choice*

Previous studies often did not find the expected benefits of causal diagrams for intervention choice. However, the designs of these studies make it difficult to draw conclusions about the current research question, namely how people use information about causal mechanisms to distinguish between several helpful interventions. In the system control studies, causal mechanisms were not relevant to the task in the first place. In the decision-making studies, the task basically required participants to search for the diagram node that represented the correct response option. Obviously, this setup cannot tell us whether causal diagrams allow people to make fine-grained distinctions between interventions that might all seem plausible at first glance but differ in the mechanisms by which they affect a problem. Moreover, although some studies varied the complexity of causal diagrams (Goode, 2011; Kleinberg & Marsh, 2023; Korshakova et al., 2023), they did not systematically vary the amount of mechanistic detail. Taken together, one limitation of these studies pertains to the interventions they used and one to the information they provided. The following sections will discuss both issues in turn.

## 2.3 How do interventions influence what information can possibly be useful?

Previous studies did not provide a systematically varied set of plausible interventions that mitigate a problem via different causal mechanisms. Therefore, knowledge about mechanisms was not actually necessary to solve the task, so there was no reason to assume it should help. To test the impact of mechanism information, interventions are needed that people cannot easily evaluate without knowing the mechanisms, and thus mechanism information can potentially make a difference. Specifically, this means that the interventions should be sufficiently unfamiliar and thus hard to distinguish, sufficiently plausible, and systematically varied in their underlying mechanisms.

One aspect of the last requirement is whether interventions treat symptoms or root causes. Focusing on symptoms rather than addressing the actual causes is a common problem in many real-world contexts such as medicine (Nogales et al., 2022). In the present study, treating root causes means to intervene early in a causal chain, removing the conditions that enable a problem to develop (e.g., in the case of oil in a car's coolant reservoir: exchanging a deformed head gasket, preventing its deformation). Conversely, treating symptoms means to merely remove the visible indicators of the problem (e.g., flushing the coolant system). However, given that the underlying causes remain, the problem will reoccur shortly (e.g., the deformed head gasket will continue to let oil leak into the cooling system). Thus, treating symptoms can be considered a quick fix, while treating root causes is a more sustainable solution. Presumably, if people were asked whether they prefer to treat root causes or symptoms, the answer would be obvious. However, in reality, the available interventions are not labelled in that way, so it is less clear which interventions people would choose.

*2.3.1 Arguments for preferring the most direct interventions*

Informally, it is widely known that people often focus on symptoms and apply quick fixes. This might not be out of sheer ignorance but because they prefer interventions that are more directly related to the problem within a causal chain. Actually, there are good reasons to prefer direct or proximal interventions (i.e., those that are closer to the effect) over less direct, distal ones (cf. Edwards et al., 2015). First, if causal links are not deterministic, distal interventions have a lower probability of success. Second, effects can dissipate as they travel through a network of causal factors. People seem to assume that such dissipation occurs as a default – even when this is not actually the case (White, 1997, 2000). Accordingly, they conclude that the effects of interventions get weaker as the number of steps between an intervention and an outcome increases, so intervening on root causes cannot guarantee that the problem actually gets solved.

*2.3.2 Arguments for preferring interventions on root causes*

Despite these reasons to prefer more direct interventions, several studies found that people actually chose to intervene on root causes (Edwards et al., 2008, 2015; Lagnado & Sloman, 2006; Yopchick & Kim, 2009). This tendency was strengthened when people aimed to achieve long-term goals (Edwards et al., 2015). Conversely, asking them to focus on short-term goals or to alleviate symptoms did not consistently make them prefer direct interventions (Edwards et al., 2015; Yopchick & Kim, 2009). Instead, they were motivated to remove the most generative factor.

*2.3.3 Summary: Characteristics of interventions*

Interventions can focus on different locations within a causal chain. There are reasons to expect either that people will mostly treat symptoms or that they will prefer to intervene on root causes. In order to notice the difference, people need to be aware of the causal mechanisms. In the present study, this was manipulated by presenting different types of causal information.

## 2.4 What features of causal information influence the choice of interventions?

A second potential influence on how causal diagrams could affect intervention choice is the type of information they provide. In principle, causal diagrams can show several aspects of causal relations. Perhaps the two most prominent ones are *whether* a variable increases or decreases another (i.e., presence and polarity of causal relations) and *how* it does so (i.e., causal mechanisms). Previous studies often provided information about presence and polarity, and sometimes even about mechanisms. However, they usually did not systematically vary features of the mechanism information such as its level of detail or its embedding in the context of other variables. Thus, to date it is not sufficiently clear what features of causal diagrams affect people's ability to distinguish between interventions. Some constraints and considerations necessary to answer this question will be discussed in this section.

*2.4.1 Presence and polarity of causal relations*

Information about the presence and polarity of causal relations may suffice especially when the specific problem content is not task-relevant (Goode, 2011; Goode & Beckmann, 2010, 2016). However, when content does matter and the domain is not well-known to a problem solver, mere presence and polarity information is of limited use. While it allows for a broad distinction between helpful, harmful, and ineffective interventions, it does not show why an intervention is more or less suitable. This can foster biases in distinguishing between multiple helpful interventions, because it opens the door for all kinds of speculations. For instance, people base their causal reasoning on their prior experiences of acting on objects (White, 2014) and on their intuitive theories of physics (Griffiths et al., 2004). Resulting assumption might be, for instance, that inanimate objects do not move without an external cause or that chemical reactions only unfold after a delay. That is, people insert non-explicit mechanisms into the causal information provided to them (Park & Sloman, 2013). Such prior knowledge and mechanistic beliefs constrain what causes they consider plausible. This can give rise to valid or invalid causal inferences. Many causal relations simply are not understandable without the intermediate mechanism steps (Johnson & Ahn, 2015). Especially in unfamiliar domains, it therefore seems useful to direct people's ideas about causal mechanisms onto the right tracks by making the mechanisms explicit.

*2.4.2 Causal mechanisms*

To make mechanisms explicit, causal diagrams can be enriched with nodes representing the mechanistic steps that mediate between an intervention and a problem. Such mechanism information should allow people to make fine-grained distinctions between interventions, because they can directly compare them and check which interventions treat the underlying root causes and which ones do not. But how do people use information about causal mechanisms? Several studies (mostly using verbal explanations rather than causal diagrams) emphasised the subjective appeal of mechanism information. People tend to prefer explanations that make causal mechanisms explicit (Vasil & Lombrozo, 2022) and such mechanistic explanations can increase their subjective ease of understanding (Fernbach et al., 2013; Vasil & Lombrozo, 2022; Zemla et al., 2023). However, knowledge about causal mechanisms also has practical consequences. Broadly speaking, it supports inference and prediction (for an overview see Johnson & Ahn, 2017). For instance, mechanistic explanations support the generalisation to new situations (Vasil & Lombrozo, 2022) and allow people to infer under which circumstances causal relations do or do not hold (Blanchard et al., 2018). Thus, explaining causal mechanisms seems like a promising way to support problem solving. However, people may also misinterpret causal mechanisms added to a causal diagram as new intervening variables that make a relation less direct (Stephan et al., 2021). This can make them judge causal relation as weaker or more uncertain.

*2.4.3 Mechanistic detail*

When providing information about causal mechanisms, the next question is how much mechanistic detail should be included. It is important to note that in principle, you can insert any number of mechanism steps into a given causal chain to explain how a cause leads to its effect (Stephan et al., 2021). However, from a cognitive perspective, the amount of mechanistic detail is not arbitrary. On the one hand, people prefer finer-grained causal explanations with more mechanistic detail (Bechlivanidis et al., 2017; Hopkins et al., 2016; Liquin & Lombrozo, 2018; Sulik et al., 2023). Such detail can increase people's subjective understanding (Fernbach et al., 2013) and eliminate their aversion to complex causal structures (Zemla et al., 2023). On the other hand, mechanistic detail may not always be useful and can even have costs. For one, it can induce a false sense of understanding (Hopkins et al., 2016; Zemla et al., 2023). Conversely, it can also make some people (especially those who are low on cognitive reflection) disprefer options as the explanations get more detailed and thus potentially overwhelming (Fernbach et al., 2013).

*2.4.4 Causal mechanisms in context*

Finally, besides adding details into a causal mechanism chain itself, this chain can also be embedded into a context of other variables. An obvious risk is that this context information is just irrelevant clutter that has no cognitive benefits or is even harmful. Partial knowledge can be sufficient to achieve high performance in problem solving tasks (Gary & Wood, 2011): It may be enough to know about key causal principles, with no need to be aware of the complete context of other variables. Moreover, irrelevant detail can impair the learning about causal mechanisms (Harp & Maslich, 2005). To assess the impact of added complexity, some studies systematically varied the amount of context variables included in causal diagrams (Kleinberg & Marsh, 2023; Korshakova et al., 2023). Higher diagram complexity impaired performance in choosing correct causes and interventions. In these studies, high complexity meant that a causal diagram had up to twelve nodes (Korshakova et al., 2023). However, even just adding one or two more nodes to a simple diagram hampered its use (Kleinberg & Marsh, 2023). Overall, people did better when only those nodes were shown that were needed for a specific decision. One reason why causal diagrams with more context variables impaired decisions might be that they made it harder to find the relevant nodes. Moreover, additional context variables and branching can also reduce people's impressions of causal strength and thus make the effectiveness of a particular chain appear less certain (Stephan et al., 2023; Stephan et al., 2021; White, 2000). Corroborating this conclusion, when more nodes and connections increased the complexity of causal diagrams, people chose to intervene on more direct and less distal causes (Kleinberg & Marsh, 2023).

*2.4.5 Summary: Characteristics of causal information*

Previous studies observed problems in using causal diagrams in general and using complex diagrams in particular. Accordingly, it has been argued that causal diagrams should be kept to a minimum, excluding variables that are not essential or cannot be changed (Kleinberg, 2024). However, such conclusions might be premature. Given that previous tasks only required participants to find the one correct node, mechanistic complexity added nothing but clutter. It is not too surprising that it is easier to find a relevant node in a diagram with two or three nodes than in a diagram with twelve nodes. Thus, the observed simplification benefits resulted from removing irrelevant information. Nothing about the more complex diagrams in these studies could possibly have helped participants to make better decisions. For instance, an aspect not investigated so far is that embedding a mechanism chain in the context of other variables could also provide structure, thereby facilitating the comparison between interventions. To investigate how mechanism information and its complexity in causal diagrams affect the choice between multiple helpful interventions, the present study manipulated features of both the interventions and the provided causal information.

## 2.5 Present study

How does information about causal mechanisms change people's intervention preferences? The present study tested the assumption that knowing the mechanisms of interventions enables people to distinguish between multiple helpful interventions and therefore choose more sustainable ones. This was investigated in a computer-based equine hoof trimming scenario. Hoof trimming is a suitable application domain as it is unfamiliar to most people but comparably easy to explain. This is because at least for current purposes, the inherent complexity of the domain can be reduced to issues of mechanics. This allows people to build on their intuitive theories of physics for causal reasoning (Griffiths et al., 2004). The more a causal relation shares features with actions on objects, the easier it is to learn and apply (White, 2014). Accordingly, people can presumably imagine the effects of interventions on a horse's hoof when the physical mechanisms are made explicit. Basically, you can remove or intentionally not remove horn material in different parts of a hoof to support a balanced growth. For instance, the overgrowth of a particular hoof structure can create leverage that tears neighbouring structures apart. The resulting fractures get infested with bacteria that cause thrush. Thus, sustainable interventions need to counteract the overgrowth. Conversely, if you only fix symptoms by disinfecting thrush, the underlying root causes are not removed and thus the problem will reappear within a few days.

The general procedure of the study was as follows. Participants were confronted with five hoof problems (e.g., thrush in the lamellar connection of the toe wall). For each problem, six interventions were available. One of them treated a root cause in a sustainable manner but still was performed relatively close to the problem, both spatially and in the causal chain (henceforth called *Direct* intervention). A second one treated a more distal root cause earlier in the causal chain and was performed far away from the problem site (*Indirect*). Conversely, a third intervention only fixed symptoms that appeared right at the problem site (*Symptom*). The fourth intervention did not affect the problem as it treated a hoof structure that was functionally independent from it (*Unrelated*) and the fifth one even aggravated it by performing the opposite action of Direct interventions (*Counterproductive*). Moreover, a sixth intervention was structurally identical to Direct interventions but had a hidden unintended consequence (*Side Effect*). However, the latter was not recognisable for participants, because the intervention was originally designed for comparison with another experiment. In that experiment, participants received information about the functions of the trimmed hoof structures and the effects of interventions on this function. In that way, they could avoid harmful side effects of an otherwise helpful intervention. However, the other experiment was aborted, because contrary to the instructions, more than half of the participants refrained from checking the function information. In the present experiment, interventions with side effects were not identifiable as such for non-experts. They still were included in the analyses as they allowed to rule out that stimulus-specific effects might be responsible for participants' evaluations of Direct interventions.

Information about the causal relation between an intervention and a problem was provided with varying levels of detail. In a baseline condition (*I*), participants only saw the intervention (e.g., shortening the toe wall from the outside) but not its effects on the problem. In this way, it could be assessed whether the interventions were perceived as more or less effective due to their inherent features (e.g., name, location, action type) rather than due to the causal information. In a second condition (*IP*), participants were informed about the presence and polarity of causal relations. That is, they learned whether the intervention blocked the problem, aggravated it, or was unrelated to it. This condition was designed to assess how ratings are affected merely by knowing about causality but without any information about causal mechanisms. A third condition (*IMP*) provided basic information about the key mechanism between the intervention and problem. Importantly, this condition did not distinguish between most interventions as all Symptom interventions shared the same key mechanism (e.g., leverage of toe wall). Seeing the key mechanism might still affect ratings of the four helpful interventions in different ways, because it made the relation understandable for the more direct interventions (i.e., Direct, Side Effect, Symptom) but not for the intervention that started early in the causal chain (i.e., Indirect). In a fourth condition (*IMMP*), more mechanistic detail was added to the causal chain (e.g., explaining step by step how shortening the toe wall prevents thrush). This allowed participants to understand how each intervention worked. However, it did not support the comparison of interventions. In contrast, the comparison of interventions was facilitated in a final condition (*IMMPC*), where each intervention was overlaid on one and the same causal diagram that included a variety of context variables (e.g., bone structure of the horse, moisture of the ground). This context might be considered irrelevant, because participants did not know the state of the additional variables (e.g., whether there were any anomalies in the horse's bone structure). However, a shared visual structure made the differences between interventions highly salient (e.g., that Symptom interventions left the entire problematic mechanism chain unchanged, while Indirect interventions already blocked the very first node of this chain).

In general, it was expected that with more detailed mechanism information, it gets easier to distinguish between multiple helpful interventions that vary in their sustainability. The following hypotheses were tested (see Figure 1 for an illustration). First, with no causal information, participants cannot reliably distinguish between interventions but preferably fix symptoms. Second, basic causal information about the presence and polarity of intervention-problem relations allows for a broad good/bad/irrelevant distinction between helpful, counterproductive, and unrelated interventions. However, it does not enable any further differentiation between the helpful interventions (i.e., Direct, Indirect, Side effect, Symptom). Still, Indirect interventions were expected to be rated somewhat lower as they were performed far away from the problem site. Third, providing only the key mechanism does not change this pattern of results. Fourth, detailed mechanism chains allow participants to make fine-grained distinctions between helpful interventions. Fifth, this ability to differentiate is further strengthened by embedding mechanism chains into the context of other variables. Taken together, it was expected that more detailed mechanism information reverses participants' preferences, making them less susceptible to symptom fixing and more appreciative of indirect interventions that address a problem's root causes. In other words, supporting the understanding and comparison of causal mechanisms should make people appreciate more sustainable interventions.

**Figure 1.** Illustration of the hypotheses. The figure groups basic causal information (IP/IMP) on the one hand and detailed mechanisms (IMMP, IMMPC) on the other. This is because within each of these groups, only gradual differences were expected instead of qualitative pattern changes, and there was no theoretical basis to make more specific predictions.

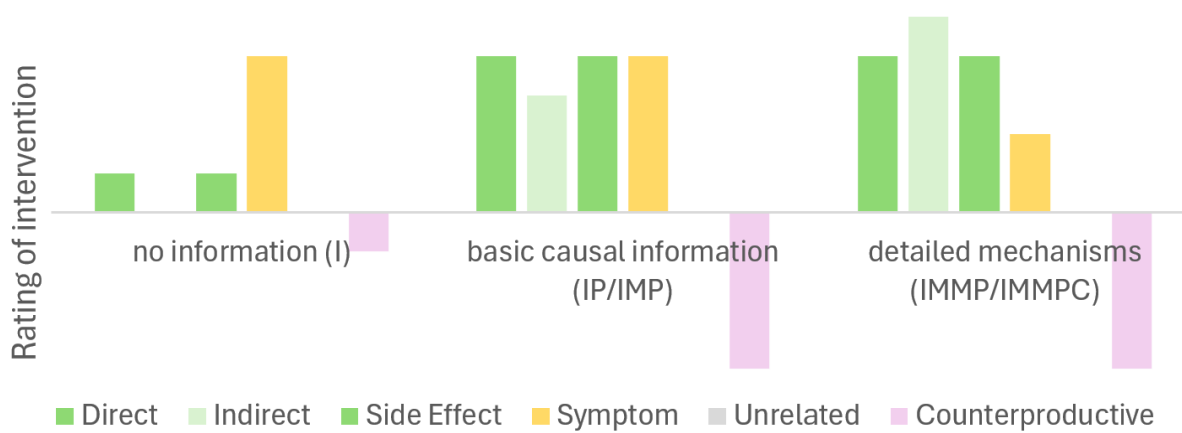

# 3. Methods

## 3.1 Open Science

All instruction materials, stimuli, raw data, data analysis scripts, and data analyses are made available via the Open Science Framework: https://osf.io/rj79b

## 3.2 Participants

Fifty-one participants were recruited via the TUD Dresden University of Technology's participant pool and up to four participants performed the experiment in parallel. Four participants were excluded for not complying with the instructions as they did not consistently check the causal information. Thus, the final sample consisted of 47 participants (32 female, 15 male) with an age range of 18 to 54 years ($M$ = 26.6, $SD$ = 7.9). All participants were native German speakers and had no prior experience with hoof trimming. They received partial course credit or a monetary compensation of 10 € per hour. Written informed consent was obtained and all procedures followed the principles of the Declaration of Helsinki.

## 3.3 Apparatus and stimuli

*3.3.1 Technical setup*

The experiments were run in a lab room on four computers (screen size 24"). A standard computer mouse was used as an input device. The experiment was programmed with the Experiment Builder (SR Research, Ontario, Canada, Version 2.6.11).

*3.3.2 Instruction video and materials*

Prior to the experiment, participants watched an instruction video of 12:36 minutes that was based on a Microsoft PowerPoint presentation with voice overlay. In the first part, information about hooves and hoof trimming was provided. The basic anatomy of a horse's hoof was explained, with all relevant parts being named and marked on a photograph. It was further explained that imbalances and shape distortions in the hoof could cause problems like fractures and thrush. This was illustrated using a 3D animation and various photographs of problematic hooves. Participants were informed that different interventions could prevent or cure such problems but that the right interventions had to be chosen. The second part of the video described the experimental task. A screenshot of the interactive overview screen was shown along with instructions on how to use it for checking the causal information and evaluating the interventions. Moreover, the five types of causal information were introduced and explained. Importantly, no information about intervention types or strategies was provided (e.g., participants were not made aware of the difference between treating symptoms versus root causes). To alleviate the need to memorise facts from the instruction video, participants had access to three printed A4 sheets during the experiment. One showed an anatomical image of a hoof with labelled parts, one showed two distorted hoof shapes (diagonal hoof and hoof with fenders), and one showed the icons used to represent each hoof shape from below, from the front, and from the side.

*3.3.3 Stimuli*

The experiment consisted of five scenarios, each featuring a different hoof problem. For each problem, six interventions systematically varied in their features. To help participants choose between these interventions, five types of causal information (one per scenario) revealed whether and how they affected the problem. All stimuli and materials were presented in German.

*Problem scenarios*. Five hoof problems had to be solved that appeared in different parts of hooves with characteristic shape distortions (i.e., long hoof, diagonal hoof, hoof with fenders). The problems were thrush in the lamellar connection of the toe wall, thrush in the right collateral sulcus of the frog, a fracture with thrush in the outer side wall, a fracture in the bearing surface of the inner side wall, and a fracture in the angle of the bar. Importantly, in each case the apparent problem (i.e., fracture and/or thrush) only represented a symptom of the imbalances or distortions in the hoof and thus sustainable interventions had to treat these underlying causes.

*Intervention types.* For each scenario, six interventions were available that differed in their effects on the hoof problem. *Direct* interventions blocked the problem in a sustainable manner by removing the imbalances or distortions that caused it (e.g., shortening a particular hoof wall). These interventions were performed near the problem site (e.g., on a fractured hoof wall, but not exactly within the fracture). Direct interventions served as a reference from which the other five intervention types differed systematically. *Indirect* interventions operated via the same mechanism chain as Direct interventions but started three steps earlier in the causal chain. Thus, they treated an even deeper root cause and therefore were at least as beneficial as Direct interventions, if not more. However, the intervention was performed on a completely different but functionally related hoof structure (e.g., side wall instead of frog, inner instead of outer side wall). Given that causal inference is inhibited when causes do not resemble their effects (White, 2009), it presumably was impossible to understand the benefits of indirect interventions without knowing the underlying causal mechanisms. *Side Effect* interventions had beneficial impacts on the problem but came with an unintended side effect. However, this side effect was not revealed to participants in the present experiment. Thus, based on the available information, the features of Side Effect interventions were identical to those of Direct interventions. *Symptom* interventions had beneficial impacts on the problem but only treated its symptoms (e.g., disinfecting or cutting out a fracture with thrush), while not addressing the underlying imbalances and distortions in the hoof. They were performed at the exact same spot where the problem appeared. *Unrelated* interventions had no impact on the problem. Just like Indirect interventions, they operated on a different hoof structure, but on a functionally unrelated one. Finally, *Counterproductive* interventions aggravated the problem by performing the opposite operation of Direct interventions (e.g., letting a hoof wall grow rather than shortening it, or vice versa). Thus, they affected the problem via the same mechanism chain as Direct interventions but promoted rather than blocked the problematic causal chain. To understand the effects of these interventions, participants had access to different types of causal information.

*Causal information conditions.* Participants received information about the causal relations between interventions and problems in the form of causal diagrams (see Figure 2). The nodes of these diagrams represented the intervention, problem, mechanisms, and context factors and were shown as coloured boxes, depending on information condition. The contents varied between six causal information conditions. In the first one, referred to as condition *I*, participants only saw the name of the intervention as a purple node. In a second condition, *IP*, the presence and polarity of a causal relation were visualised. By connecting the purple intervention node to a red problem node, it was indicted whether the intervention aggravated the problem (solid line with arrowhead), blocked it (dashed line that ended in a circle), or had no effect on it (no link). The third condition, *IMP*, additionally provided the key mechanism that mediated between the intervention and problem. This mechanism was shown in a single yellow node that was identical for the Direct, Indirect, Side Effect, and Counterproductive interventions, and only differed for Symptom interventions (for Unrelated interventions, no mechanism existed). In a fourth condition, *IMMP*, the causal information was split up into a chain of mechanism steps, with their number depending on intervention type (i.e., 4-5 for Direct, Side Effect, and Counterproductive, 7-8 for Indirect, 1 for Symptom, 0 for Unrelated). Finally, in condition *IMMPC*, this mechanism chain was embedded in context information about other variables that entered the mechanism chain between the intervention and problem. The same context – and thus the same structure of the causal diagram – was shown for each intervention as a network of influences. Within this network, the mechanism chain of the currently selected intervention was highlighted in yellow, while all other influences (including the mechanisms of other interventions) were presented as white context nodes. Accordingly, given that the full diagrams consisted of 18-20 nodes, the number of context nodes depended on intervention type (i.e., 11-14 for Direct, Side Effect, and Counterproductive, 8-11 for Indirect, 15-17 for Symptom, 16-18 for Unrelated). Importantly, the context information was not necessary to understand the effects of an intervention. However, by highlighting the respective mechanism chains in the same visual structure, it presumably facilitated the comparison of interventions.

**Figure 2.** Causal information conditions for each intervention type. (A) Intervention alone (I). (B) Causal relation between intervention and problem (IP). (C) Key mechanism mediating between the intervention and problem (IMP). (D) Detailed mechanism chain mediating between the intervention and problem (IMMP). (E) Detailed mechanism chain embedded in a broader causal context (IMMPC), exemplarily shown for Indirect and Symptom interventions.

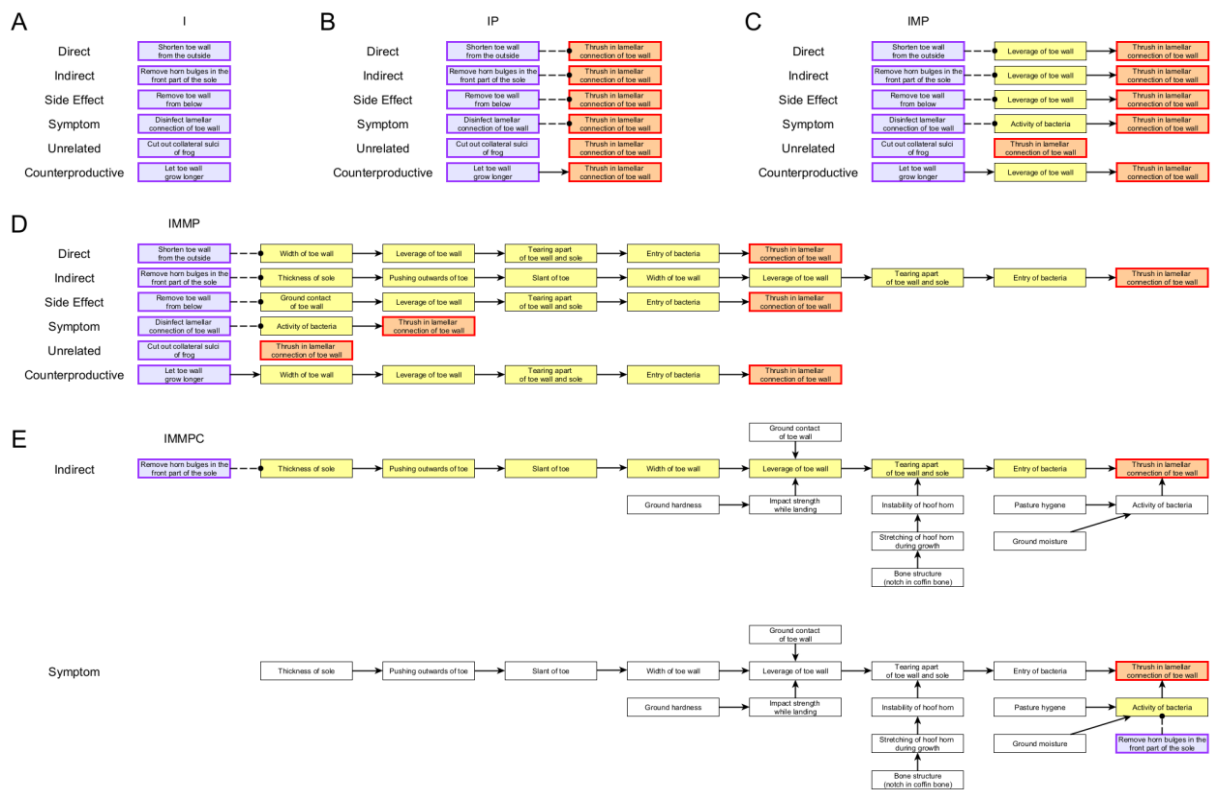

*Screens.* In each trial, three types of screens were shown: an instruction screen, six causal information screens, and an interactive overview screen. Instruction screens indicated which type of causal information would be provided in the following trial and reminded participants of its meaning. Causal information screens visualised the causal diagram for the currently checked intervention. An interactive overview screen (see Figure 3) stated the problem, listed the available interventions, provided a rating scale for each of them, and contained buttons for navigation. The problem label was presented in the top row, together with a schematic drawing of the problem as a red marking on a hoof icon (viewed from below, from the front, or from the side). Interventions were listed at the left-hand side of the screen, each accompanied by a hoof icon with a purple marking that showed where the intervention was performed and what it did to the hoof. Moreover, for each intervention there was a button labelled "How does the intervention affect the problem?" (leading participants to the respective causal information screen). At the right-hand side, a rating scale was available for each intervention. The scales were titled "How suitable is the intervention for solving the problem?" and ranged from -5 (very poorly) over 0 (neutral) to 5 (very well). Upon clicking a scale position, it was filled with a purple dot. Upon clicking an intervention label to choose the intervention as the best one, it was surrounded by a purple frame. Finally, at the bottom-right corner of the screen there was a button titled "Continue to next trial".

**Figure 3.** Stimulus example showing an interactive overview screen.

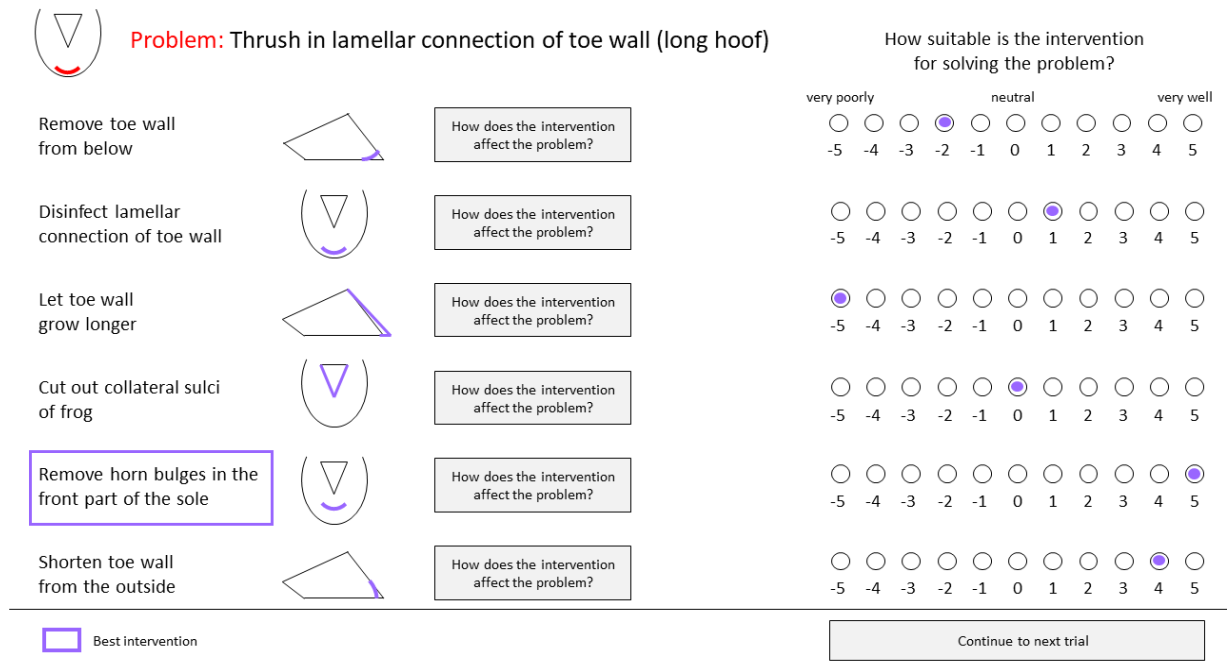

### 3.4 Procedure

After providing written informed consent and demographic information, participants watched the instruction video and then performed the experiment. The experiment used a 5 (*causal information: I, IP, IMP, IMMP, IMMPC*) × 6 (*intervention type: Direct, Indirect, Side Effect, Symptom, Unrelated, Counterproductive*) within-subjects design. It consisted of five trials, each corresponding to a different problem scenario. Between scenarios, the type of causal information was varied, with the assignment being counterbalanced across participants so that each scenario appeared about equally often in each causal information condition. Moreover, the order of causal information conditions was randomised, with the constraint that each appeared about equally often at each sequential trial position. Within a trial, participants first saw the instruction screen and by clicking the "Start trial" button, they could move on to the interactive overview screen. On that screen, they were free to investigate and rate the interventions in any order they liked, and also to flexibly revise their ratings as often as they wanted. To check the causal information associated with an intervention, they had to click the respective button, which brought them to the causal information screen. Once they were finished checking the causal information, they could return to the overview screen by clicking a "Back to overview" button. Finally, participants had to choose the best intervention by clicking its label, which made a purple frame appear around this label. Once participants were finished, they could move to the next trial by clicking the "Continue to next trial" button. The study took about half an hour.

## 3.5 Data analysis

Two dependent variables were analysed: participants' ratings of the interventions and their choices of which one was best. To assess how causal information affected the ratings, they were analysed using a 5 (*causal information: I, IP, IMP, IMMP, IMMPC*) × 6 (*intervention type: Direct, Indirect, Side Effect, Symptom, Unrelated, Counterproductive*) repeated-measures ANOVA. If sphericity was violated, a Greenhouse-Geisser correction was applied and the degrees of freedom were adjusted accordingly. To determine statistical significance, an alpha level of *p* < .05 was used and all pairwise comparisons were performed with Bonferroni correction. To assess which intervention participants considered the best in each causal information condition, the number of choices was analysed descriptively.

## 4. Results

### 4.1 Ratings of interventions

Participants' ratings of the six interventions were compared between the causal information conditions. The ANOVA revealed main effects of causal information, $F(2.3, 105.0) = 30.327$, $p < .001$, $\eta_p^2 = .397$, and intervention type, $F(4.1, 187.9) = 507.683$, $p < .001$, $\eta_p^2 = .917$, as well as an interaction of both factors, $F(6.7, 307.8) = 27.719$, $p < .001$, $\eta_p^2 = .376$ (see Figure 4A).

**Figure 4.** Participants' ratings and choices of interventions, depending on causal information condition and intervention type. (A) Ratings of interventions. (B) Choices of the best intervention. Error bars represent standard errors of the mean.

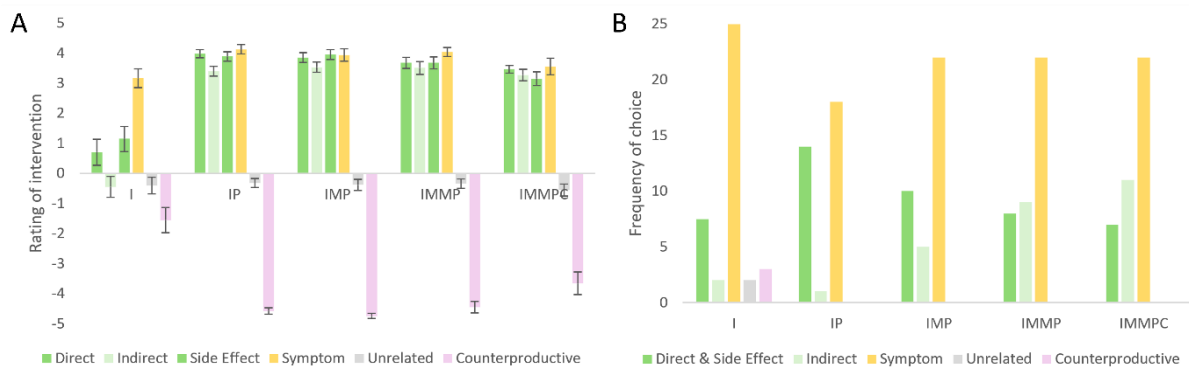

Pairwise comparisons indicated that ratings were lower with no information (i.e., I) than with any type of causal information (i.e., IP, IMP, IMMP, and IMMPC), all *p*s < .001, while there were no differences between the latter, all *p*s > .413. However, an exploratory analysis partly revised this conclusion. Figure 4A suggests that overall, more complex mechanism information dampens participants' ratings. Thus, there might be differences between the causal information conditions after all, which cannot be revealed by the ANOVA because such dampening works in opposite directions for interventions that are rated positively versus negatively. As a result, the effects might cancel each other out. Therefore, the ANOVA was repeated with only the four helpful interventions (i.e., Direct, Indirect, Side Effect, Symptom). Within this reduced ANOVA, interventions were rated lower with IMMCP information (3.4) than with IP information (3.9), $p = .001$, and IMP information (3.8), $p = .014$. Only the difference to IMMP information (3.7) was non-significant, $p = .096$. Thus, helpful interventions seemed to be perceived as less helpful when embedded in the context of other variables.

Returning to the main ANOVA, pairwise comparisons between intervention types indicated that ratings were lower for Counterproductive interventions (-3.8) than for all other interventions, all *p*s < .001, and lower for Unrelated interventions (-0.4) than for all helpful interventions, all *p*s < .001. Among the helpful interventions, Symptom interventions (3.8) were rated higher than the three interventions that treated root causes, all *p*s < .031. Moreover, Direct interventions (3.1) and Side Effect interventions (3.2) were rated higher than Indirect interventions (2.7), both *p*s < .02, while not differing from each other, *p* > .9.

The differences between intervention types also depended on causal information (see Table 1 for an overview of all means and standard deviations). First, with no information (I), Symptom interventions were rated higher than all others, all *p*s < .001. Additionally, participants were able to distinguish some of the helpful interventions from the non-helpful ones: Direct interventions were rated higher than Counterproductive interventions, *p* = .036, and Side Effect interventions were rated higher than Counterproductive and Unrelated interventions, both *p*s < .04. Conversely, Indirect interventions did not differ from the two non-helpful interventions, both *p*s > .9.

Second, whenever any causal information was available (regardless of its type), this created a clear good/bad/irrelevant trichotomy between intervention types: Ratings were considerably higher for helpful interventions than Unrelated and Counterproductive interventions, all *p*s < .001, and lower for Counterproductive interventions than Unrelated ones, all *p*s < .001.

Third, subtle differences between the four helpful interventions depended on the complexity of causal information. Whenever any causal information was available, Symptom interventions were no longer superior to Direct and Side Effect interventions, all *p*s > .9. Moreover, when only the presence and polarity of causal relations were revealed (IP), all three interventions with more direct effects (i.e., Direct, Side Effect, Symptom) were rated higher than Indirect interventions, all ps < .02. This inferiority of Indirect interventions was reduced when information about the key mechanism was added (IMP). Now they were only rated minimally lower than Side Effect interventions, *p* = .043, but no longer differed from Direct and Symptom interventions, both *p*s > .3. Finally, when detailed mechanism information was provided (IMMP and IMMPC), all differences between helpful interventions disappeared, all *p*s > .8.

Table 1. Means and standard deviations (in parentheses) of participants' ratings of interventions as well as best intervention choice frequencies and percentages (in parentheses), depending on causal information and intervention type.

|  |  | I | IP | IMP | IMMP | IMMPC |
|---|---|---|---|---|---|---|
| Ratings of interventions | Direct | 0.7 (3.0) | 4.0 (0.9) | 3.9 (1.1) | 3.7 (1.3) | 3.5 (0.9) |
|  | Indirect | -0.4 (2.4) | 3.4 (1.1) | 3.5 (1.2) | 3.5 (1.5) | 3.3 (1.3) |
|  | Side Effect | 1.1 (2.9) | 3.9 (1.1) | 4.0 (1.1) | 3.7 (1.4) | 3.1 (1.6) |
|  | Symptom | 3.2 (2.2) | 4.1 (1.1) | 3.9 (1.4) | 4.0 (1.0) | 3.6 (1.9) |
|  | Unrelated | -0.4 (1.9) | -0.3 (1.0) | -0.4 (1.2) | -0.3 (1.1) | -0.6 (1.3) |
|  | Counterproductive | -1.6 (2.9) | -4.6 (0.7) | -4.7 (0.6) | -4.4 (1.3) | -3.7 (2.6) |
| Choices of best intervention | Direct | 8 (17.0 %) | 12 (25.5 %) | 8 (17.0 %) | 5 (10.6 %) | 11 (23.4 %) |
|  | Indirect | 2 (4.3 %) | 1 (2.1 %) | 5 (10.6 %) | 9 (19.1 %) | 11 (23.4 %) |
|  | Side Effect | 7 (14.9 %) | 16 (34.0 %) | 12 (25.5 %) | 11 (23.4 %) | 3 (6.4 %) |
|  | Symptom | 25 (53.2 %) | 18 (38.3 %) | 22 (46.8 %) | 22 (46.8 %) | 22 (46.8 %) |
|  | Unrelated | 2 (4.3 %) | 0 (0.0 %) | 0 (0.0 %) | 0 (0.0 %) | 0 (0.0 %) |
|  | Counterproductive | 3 (6.4 %) | 0 (0.0 %) | 0 (0.0 %) | 0 (0.0 %) | 0 (0.0 %) |

Taken together, being informed about the mere presence and polarity of a causal relation enabled participants to clearly distinguish helpful interventions from harmful and irrelevant ones. It also reduced participants' tendency to rate symptom fixing as superior to interventions that treated the root causes of a problem. However, only with more detailed information about causal mechanisms, participants came to appreciate direct and indirect interventions as equally helpful.

It is important to note, however, that this change in rating patterns between the four helpful interventions did not come about as expected. When looking back at the visual illustration of the hypotheses in Figure 1, it had been assumed that detailed mechanism information decreases the ratings for Symptom interventions and increases them for Indirect interventions, while the more direct ways of treating root causes (Direct and Side Effect) are less affected. In fact, the opposite was found, as revealed by pairwise comparisons of each intervention type's ratings between the causal information conditions. Whereas the ratings for Symptom and Indirect interventions did not change when mechanistic detail was added, all $ps > .9$, ratings decreased from IP to IMMPC for Direct, $p = .019$, and Side Effect interventions, $p = .027$.

## 4.2 Choices of best intervention

Participants' intervention choices were compared between the causal information conditions. For this purpose, the values for Direct and Side Effect interventions were averaged, because based on the information available in the present experiment, they were indistinguishable for novice participants. This also was corroborated by the rating results. Thus, if their choice frequencies were analysed individually, two equivalent options would compete with each other, hampering the interpretation of the results. However, the original, separate values can be found in Table 1.

When comparing the choice frequencies (see Figure 4B), the most notable observation was that Symptom interventions were chosen most frequently by far across all causal information conditions. This was most obvious when no information was available (I). However, even with detailed mechanism information and its embedding in the context of other variables, Symptom interventions were chosen by almost half of the participants (even with the exact same number of choices in IMP, IMMP, and IMMPC). While interventions that treated root causes never caught up with Symptom interventions, there was a notable shift between them. Indirect interventions were highly dispreferred without information (I) or with only basic causal information (IP). However, they got a boost when the key mechanism was revealed (IMP), caught up with the more direct interventions (i.e., Direct & Side Effect) when detailed mechanisms were added (IMMP), and even surpassed them when the mechanism chain was embedded in the context of other variables (IMMPC).

## 5. Discussion

Successful problem solving requires decision-makers to understand how their interventions affect the problem at hand. The present study asked how information about causal mechanisms shifts people's preferences between multiple helpful interventions. It was hypothesised that mechanism information helps them to be less focused on fixing symptoms and more inclined to treat the underlying root causes. The results painted a mixed picture. On the one hand, mechanism information helped people appreciate indirect interventions that treated root causes early in the mechanism chain. These highly sustainable interventions were considered inferior when causal information was either lacking or when it only showed which interventions mitigated the problem. With detailed mechanism information, indirect interventions were no longer dispreferred. Thus, elaborate mechanistic explanations did have some benefits. On the other hand, the results do not look very promising with regard to symptom fixing. A strong preference for these interventions was somewhat mitigated by causal information in general and mechanistic detail in particular. However, across all information conditions, participants still deemed symptom fixing the best intervention. That is, even when they received elaborate explanations of how distortions in the hoof led to the observed problem, they chose to ignore this entire problematic mechanism chain, leave the distortions as they are, and indicate that the best solution was to disinfect the problem site. This immunity of symptom fixing to contrary evidence is concerning and suggests that the allure of quick fixes cannot be overcome simply by informing people about causal mechanisms. The following sections will explore the potential origins of these findings, addressing issues of information complexity and discussing why the effects of mechanism information might have remained rather small.

## 5.1 The inclination to fix symptoms is highly robust

Perhaps the most striking outcome of this study was the popularity and robustness of symptom fixing. While this had been expected in the absence of causal information, it persisted even when the detailed mechanisms were revealed. To account for this unexpected finding, three explanations will consider the role of recognisability, similarity, and uncertainty.

First, participants might not have recognised symptom fixing as such. Indeed, in the post-experimental debriefing, some were quite surprised that they had not noticed it. This suggests that one way of mitigating the tendency to go for quick fixes is to emphasise the sustainability of interventions. A previous study suggests that this can indeed change intervention preferences: When participants' focus was implicitly or explicitly directed to long-term goals, they more often intervened on root causes (Edwards et al., 2015). However, numerous accounts of symptom fixing in real life make it unlikely that it can be eliminated simply by telling people to choose sustainable interventions.

A second explanation focuses on similarity. Symptom interventions were most similar to the problem, both spatially and phenomenologically. For one, they operated exactly at the problem site, while direct interventions were somewhat further removed and indirect interventions were performed on a completely different hoof structure. Participants might have been less inclined to believe that spatially distant interventions can be effective. Also, the phenomenological similarity to the problem was higher for symptom interventions. Thrush is a biochemical phenomenon, presumably leading participants to favour chemical interventions like disinfection. Generally, it is harder for people to recognise causal relations when the causes and effects are dissimilar (White, 2009) or when they stem from different mechanism domains (Dündar-Coecke et al., 2022). In line with this, people aim to match intervention types to problem types, for instance preferring medication when the problem is considered biological but psychotherapy when it is considered psychological (Yopchick & Kim, 2009).

The third explanation holds that the positive effects of treating symptoms might have seemed more certain. For one, the mechanism chains of symptom interventions were much shorter, only consisting of one step. This avoids the risk of the effects fading out along the mechanism chain (Edwards et al., 2015; White, 1997). Another aspect of uncertainty is whether a particular state is really present and problematic. When the problem description refers to thrush, there is no doubt that there is thrush and that this is a bad thing. Conversely, there was no explicit information ensuring that the undesirable steps of the other interventions' mechanism chains were indeed present (e.g., perhaps the toe wall was not really too long) and that they were indeed problematic (e.g., perhaps a long toe wall did not really cause leverage). This explanation resonates with the overall difficulty of applying general causal information on a type level to the token level of a current situation (Stephan & Waldmann, 2022). Only because a mechanism generally exists, this does not guarantee that it was in operation in a particular situation. Future studies should explore ways to support the instantiation of mechanism information.

These explanations still do not answer the question why a robust preference was observed in the present study, while some previous studies found preferences for treating root causes (Edwards et al., 2008, 2015; Lagnado & Sloman, 2006; Yopchick & Kim, 2009). In one study, participants chose to intervene on root causes even when they were explicitly asked to alleviate the symptom (Yopchick & Kim, 2009). It is not fully clear what caused these contradictory findings. One possibility is that in the present study, interventions were more difficult to compare. Previous studies typically presented a causal chain and then participants could decide which node to intervene on. Conversely, in the present study, the interventions appeared as a list of qualitatively different options. However, this does not explain why symptom interventions did not drop in popularity even when all interventions were overlaid on the same causal structure. Thus, it will be left as an open question for future studies how preferences change depending on how easy it is to compare different interventions.

## 5.2 The effects of mechanism information were small and unexpected

The effects of mechanism information on participants' ratings and choices were much smaller than expected. A look at Figure 4A reveals that the main difference between the causal information conditions emerged between no information and any causal information. Conversely, the differences between the four causal conditions were small, even though they were statistically robust. That is, adding information about causal mechanisms only had minor impacts on participants' ratings and choices. This resonates with previous findings that mechanism information is not always used. When people freely generated questions to ask whatever they wanted to know about a causal relation (either in the internet or in the lab), they rarely asked about mechanisms (Hagmayer & Engelmann, 2020). Similarly, when people could actively access the mechanism steps mediating between a cause and an effect, many did not do so, or only investigated the initial steps and then stopped (Czarnowski & Marsh, 2021). Given that people often do not actively seek information about mechanisms, they might also process it rather superficially when it is provided to them (Stephan & Waldmann, 2022). Thus, the small size of mechanism effects might have resulted from a limited processing depth. This might have been exacerbated by the rather formal presentation format of causal diagrams. It is conceivable that this led participants to predominantly process these diagrams syntactically, instead of deeply processing their contents. Informal evidence stems from participants' verbal reports about their strategies during debriefing. Some indicated that first and foremost, they had checked whether there was an arrow or dotted line with a blocker (indicating a helpful or harmful intervention). Others said that with detailed mechanism information, they preferably chose the intervention that only had one step, because it is most directly associated with the problem. While such superficial processing seems like a serious risk of causal diagrams, some studies compared causal diagrams with causal text and found similar problems with both presentation formats (Kleinberg et al., 2025; White, 1999; Zheng et al., 2020). However, the two presentation formats can induce different reasoning strategies with different impacts on intervention choice (Kleinberg et al., 2025). Thus, future studies should investigate how they affect the way people use information about causal mechanisms.

Other than the effects being small, their specific pattern was unexpected. Contrary to the hypotheses, adding mechanism information did not make participants distinguish more clearly between different types of helpful interventions. In fact, the opposite was found. Embedding mechanism chains in the context of other variables flattened out the differences between the four helpful interventions. Moreover, it dampened the ratings overall, making interventions seem less effective in general. One reason might be that participants were overwhelmed by the sheer amount of information. Indeed, detailed mechanistic explanations can be too much for some people, especially when they are low in cognitive reflection (Fernbach et al., 2013). A second reason might be that participants became more aware of other influences on the mechanism chain and the problem (e.g., bone structure, ground conditions, pasture hygiene). Thus, interventions in general might have been considered less effective. This is in line with previous findings that context variables and branching can make people perceive causal relations as weaker or more uncertain (Stephan et al., 2023; Stephan et al., 2021; White, 2000). A third and completely different explanation is that the context information might indeed have worked as intended, but with unexpected side effects. Apparently, it made participants realise that indirect interventions were beneficial after all, as indicated by an increase in their choices. This, in turn, might have reduced participants' positive opinions about direct interventions that started somewhere in the middle of a mechanism chain. The latter is in line with the observation that people least prefer causes in the middle of a causal chain (Edwards et al., 2008, 2015). Given that mechanism information enables them to see where in a causal chain an intervention is operating, it allows them to distinguish between deeper and intermediate root causes. This might have led to a decrease in the latter's ratings.

An important question is why more complex causal diagrams did have some benefits in the present study, while other studies only reported disadvantages (Kleinberg & Marsh, 2023; Korshakova et al., 2023). Several differences between the experimental setups might account for this divergence. A first one is whether interventions become more distinguishable with more complex causal information. In the previous studies, the alternatives to the correct interventions were highly implausible (e.g., watching TV or eating pizza to improve one's health). Presumably, no detailed information is needed to distinguish them from a helpful intervention. Conversely, the present study used several helpful interventions from an unfamiliar domain that were hard to evaluate without causal information. If anything, not having access to causal information gave participants a wrong idea about their relative effectiveness (i.e., highest ratings for symptom and lowest for indirect interventions). Thus, causal information in general and mechanism information in particular were actually needed in the present study, although the results suggest that participants did not make optimal use of this information. The present study also differs from previous ones in that complex information contributed more than just clutter that makes it harder to spot the relevant nodes. Moreover, in the present study, spotting the relevant nodes was not an issue as they were visually highlighted, which was previously found to eliminate disadvantages of more complex causal diagrams (Kleinberg & Marsh, 2023).

### 5.3 Limitations and future research

Several limitations of the present study relate to characteristics of both the experimental materials and the participants, raising issues of internal and external validity. For one, it is not clear how participants represented the task of evaluating the interventions. This seems particularly relevant for symptom interventions. One may argue that disinfecting thrush is not only helpful but even necessary, just like flushing the cooling system in the car example of the Introduction. It will not solve the problem, or not for long, but still it needs to be done. Accordingly, some participants might have concluded that symptom interventions should receive high ratings, despite knowing that they should not be performed in isolation. Thus, future studies might use tasks that allow for a fine-grained differentiation between what participants consider necessary versus sustainable. Another limitation of the present interventions is that they were presented as binary options (i.e., performing the intervention or not). Conversely, for most real-life interventions, good or bad is a matter of dose. In fact, it is not possible to say whether thinning out a hoof wall is good or bad, because this also depends on how much material you take away.

An obvious limitation is that participants were lacking domain expertise. While this is a general issue in psychological research, it might have had specific implications in the present study. First, it is hard to imagine that experts would have expressed a similar preference for symptom fixing. With increasing expertise, people learn more about the causal relations in a domain and become less reliant on the heuristic that valid causes resemble their effects (White, 2009). Presumably, this would also allow them to appreciate interventions that are spatially, phenomenologically, or in other ways dissimilar to the problem. That said, being an expert by no means guarantees that people agree on the effectiveness of interventions or on their causal models about them (de Kwaadsteniet & Hagmayer, 2018; de Kwaadsteniet et al., 2010). Moreover, expertise might also affect how people deal with causal diagrams, although the empirical findings on this are mixed. While there are studies reporting that causal diagrams are more helpful for people without domain experience (Zheng et al., 2020), others found that they helped experts and novices alike (de Kwaadsteniet et al., 2013). Presumably, this is largely dependent on the specific contents of the diagrams and the task they are to support.

A related limitation concerns the role of interindividual differences, which were not systematically investigated although they were clearly present. For instance, some participants reported that symptom interventions were insufficient and thus gave very low ratings to them. Similar differences were informally observed in the way people used the causal information. This fits with the findings that people differ in whether they appreciate mechanistic detail (Fernbach et al., 2013) and what strategies they use to deal with information about causal mechanisms (Stephan & Waldmann, 2022).

There are many exciting prospects for future research. From a practical perspective, an important question is how to support people in adequately using information about causal mechanisms. A critical issue seems to be that they process the information superficially, for instance by focusing on syntactic features of causal diagrams. Therefore, effective information transmission should foster a deeper engagement with causal mechanisms. One way to do so is prompting people to actively construct causal diagrams. This can enhance causal learning and understanding (Easterday et al., 2009; Jeong & Lee, 2012), but it does not guarantee that people also get better at solving problems (Öllinger et al., 2015). Thus, an alternative way to support the active engagement with causal mechanisms is to combine predictive and diagnostic reasoning. The present study only provided predictive causal information by specifying the effects of interventions. However, it is known that people are more likely to neglect alternatives in predictive than diagnostic tasks (Fernbach et al., 2011). Moreover, they choose better interventions after being trained to diagnose than to predict the state of a complex system (Schoppek, 2004). In a similar vein, the biases observed in the present study might be mitigated by prompting people to elaborate how a problem has come about, for instance by generating possible causes and explaining their mechanisms with the help of a causal diagram. It seems likely that after having made the causes explicit, they would no longer consider symptom fixing the best intervention.

Aside from the role of mechanism information in solving a current problem, future studies could also investigate its role in the transfer of interventions to new situations. Causal diagrams might support this as they have previously been shown to improve the transfer of problem solving abilities (Putz-Osterloh, 1993). Moreover, explanations of mechanisms allow for a broader understanding and generalisation to novel circumstances (Vasil & Lombrozo, 2022). This might enable people to choose interventions based on similar mechanistic principles. For instance, when they have understood that a problem caused by leverage of a hoof structure can be fixed by reducing the levers, they might generalise this across problem instances with different low-level features. However, people's willingness to transfer knowledge about the effects of interventions depends on perceived similarity (Czarnowski & Marsh, 2022). Accordingly, transfer can only be successful and efficient if people develop reusable mental abstractions or ways of interpreting situations (Nagy et al., 2025). To support these cognitive activities and make similar mechanisms visible and assessable, information about them should systematically vary in its level of abstraction and allow people to navigate between these levels. We are currently working on modelling methods to provide such information.

Finally, it would be intriguing to study how general or domain-specific the link between mechanism knowledge and intervention choice actually is. On the one hand, it has been suggested that the effects of causal diagrams are similar across domains (Korshakova et al., 2023). However, this might depend on whether domains are mainly used as a cover story or the problem features and causal relations are fundamentally domain-specific. Domain characteristics shape cognitive work requirements (Müller & Oehm, 2019; Schmidt & Müller, 2023). For instance, in some domains, the dissipation of causal strength along causal chains or across branches of causal networks might be more common than in others (White, 1997). This, in turn, might affect whether more direct or indirect interventions are preferable. Also, causal reasoning relies on the processing of domain-specific information (Hauptman & Bedny, 2024), which is likely to affect intervention choices. For instance, people more strongly prefer to intervene on root causes in some domains than in others (Edwards et al., 2015). In consequence, the perceived value of causal information might also depend on the domain, making people consider mechanistic explanations more helpful for reasoning about some systems than others (Johnson & Keil, 2017). Thus, domain considerations provide ample exciting opportunities for future research.

## 5.4 Conclusion

At the outset of this study, it seemed almost trivial that when people understand how and why their interventions affect a problem, they can choose better ones. However, the results revealed that it is not that simple. It seems like people are highly inclined to fix symptoms, even when detailed information about causal mechanisms is telling them that other interventions are far more sustainable. Thus, in order to support a thorough consideration of qualitatively different interventions, it is not sufficient to provide causal information. Future research should investigate what is needed to help people engage in critical thinking about the causal mechanisms that make interventions sustainable.

## Acknowledgments

I want to thank Konstanze Rasch for the valuable exchange that inspired the conceptualisation of this study, as well as Judith Schmidt and Paul Weber for fruitful discussions about the research questions and methods. This work was supported by a grant of the German Research Foundation (DFG, MU 3749/3-1).## References

Ahn, W. K., Kalish, C. W., Medin, D. L., & Gelman, S. A. (1995). The role of covariation versus mechanism information in causal attribution. *Cognition*, *54*(3), 299-352. https://doi.org/10.1016/0010-0277(94)00640-7

Bechlivanidis, C., Lagnado, D. A., Zemla, J. C., & Sloman, S. A. (2017). Concreteness and abstraction in everyday explanation. *Psychonomic Bulletin & Review*, *24*(5), 1451-1464. https://doi.org/10.3758/s13423-017-1299-3

Blanchard, T., Vasilyeva, N., & Lombrozo, T. (2018). Stability, breadth and guidance. *Philosophical Studies*, *175*(9), 2263-2283. https://doi.org/10.1007/s11098-017-0958-6

Czarnowski, D. W., & Marsh, J. E. (2022). Transferring novel causal knowledge. *In* Proceedings of the Annual Meeting of the Cognitive Science Society (pp. 122-128).

Czarnowski, D. W., & Marsh, J. K. (2021). Searching for the cause: Search behavior in explanation of causal chains. *In* Proceedings of the Annual Meeting of the Cognitive Science Society (pp. 646-652).

de Kwaadsteniet, L., & Hagmayer, Y. (2018). Clinicians' personal theories of developmental disorders explain their judgments of effectiveness of interventions. *Clinical Psychological Science*, *6*(2), 228-242. https://doi.org/10.1177/2167702617712270

de Kwaadsteniet, L., Hagmayer, Y., Krol, N. P. C. M., & Witteman, C. L. M. (2010). Causal client models in selecting effective interventions: A cognitive mapping study. *Psychological Assessment*, *22*(3), 581-592. https://doi.org/10.1037/a0019696

de Kwaadsteniet, L., Kim, N. S., & Yopchick, J. E. (2013). How do practising clinicians and students apply newly learned causal information about mental disorders? *Journal of Evaluation in Clinical Practice*, *19*(1), 112-117. https://doi.org/10.1111/j.1365-2753.2011.01781.x

Dündar-Coecke, S., Goldin, G., & Sloman, S. A. (2022). Causal reasoning without mechanism. *PLoS ONE*, *17*(5), e0268219. https://doi.org/10.1371/journal.pone.0268219

Easterday, M. W., Aleven, V., Scheines, R., & Carver, S. M. (2009). Constructing causal diagrams to learn deliberation. *International Journal of Artificial Intelligence in Education*, *19*(4), 425-445.

Edwards, B. J., Burnett, R. C., & Keil, F. C. (2008). Structural determinants of interventions on causal systems. *In* Proceedings of the 30th Annual Meeting of the Cognitive Science Society (pp. 1138-1143).

Edwards, B. J., Burnett, R. C., & Keil, F. C. (2015). Effects of causal structure on decisions about where to intervene on causal systems. *Cognitive Science*, *39*(8), 1912-1924. https://doi.org/10.1111/cogs.1221021